\documentclass[twocolumn,prl,floatfix,showpacs]{revtex4}
\usepackage{epsfig}
\begin{document}
\title{Symbolic dynamics of biological feedback networks}
\author{Simone Pigolotti, Sandeep Krishna and Mogens H. Jensen}
\affiliation{Niels Bohr Institute and Niels Bohr
International Academy, Blegdamsvej 17, DK-2100 Copenhagen, Denmark} 
\homepage{http://cmol.nbi.dk}
\date{\today}
\begin{abstract}

  We formulate general rules for a coarse-graining of the dynamics,
  which we term `symbolic dynamics', of feedback networks with
  monotone interactions, such as most biological modules. Networks
  which are more complex than simple cyclic structures can exhibit
  multiple different symbolic dynamics.  Nevertheless, we show several
  examples where the symbolic dynamics is dominated by a single
  pattern that is very robust to changes in parameters and is
  consistent with the dynamics being dictated by a single feedback
  loop. Our analysis provides a method for extracting these dominant
  loops from short time series, even if they only show transient
  trajectories.
\end{abstract}
%lowdim chaos, oscillations in chemical systems, time series analysis
\maketitle 

Many biological systems can be described by directed networks, where
nodes represent different components and arrows represent
interactions. In cell biology, nodes are molecules, while arrows stand
for complex formation, protein modification, transcription regulation,
etc. Ecosystems constitute another example, where nodes are species
and arrows represent predation, competition and symbiosis. Biological
functions are often performed by specific small subnetworks, or {\em
  modules} \cite{Hartwell}.  A dynamical model of a module requires,
beyond the knowledge of the network structure, some hypothesis on the
form of the interactions, which are often poorly characterized. It is
then crucial to develop techniques to study the qualitative dynamics
of modules assuming limited information.

We present a method to obtain information about patterns in the
dynamics of biological feedback networks. Given the network structure,
i.e., which nodes activate/repress which other nodes, it is possible
to predict the ordering of maxima and minima of the dynamical
variables. Vice versa, from an experimentally observed ordering one
can obtain some information about the structure of the network.  The
method is a generalization of the one introduced in \cite{patterns}
for the dynamics of a single negative feedback loop, without any cross
links, where a unique pattern is allowed.  A More complex network
structure \cite{mallet,bechhoefer,rand,kaufman,tiana,ross} allows
multiple dynamical patterns. Moreover, a particular observed pattern
could originate from different network structures.  Nevertheless, our
method can reduce the possibilites and provides non-trivial
information that can guide experiments. Our technique also applies
when the dynamics is transient, so that information can be
obtained, for example, by watching how the concentrations of proteins
and genes belonging to a given module relax to a stationary state
after a perturbation.  This extends the use of our formalism to cases
in which oscillations are damped \cite{tiana}.

We consider a system described by $N$ dynamical variables, $x_i$,
$i=1\ldots N$, which we call ``densities'' and could represent, for
example, the concentrations of the chemical species composing the
network/module.  We assume that they evolve with time in a
deterministic way, according to a system of differential equations:

\begin{equation}
\frac{dx_i}{dt}=g_i(x_1,x_2\ldots x_N)\qquad i=1\ldots N.
\end{equation}

Many possible dynamical systems may correspond to a given network. The
only constraints we impose are that the interactions be monotonic,
i.e., each off-diagonal element of the Jacobian matrix,
$\partial_{x_j}g_i$, is either positive everywhere in phase space
(when node $j$ activates $i$), or negative everywhere ($j$ represses
$i$), or zero everywhere (no arrow from $j$ to $i$). In words,
activators are always activators and repressors are always
repressors. Indeed, transcription factors rarely switch from being
activators of a particular gene to repressors at different densities;
a predator of a particular species does not become its prey when their
abundances change. We do not require monotone self-interactions: a
variable may activate or repress itself depending on the densities.

We associate to each state $(x_1,x_2\ldots x_N)$ a {\em symbol} such
as $(+,-,-,\ldots +)$. This $N$-component sign vector describes which
densities are increasing and which are decreasing at a given time: the
$i$-th component is just the sign of $g_i(x_1,x_2\ldots x_N)$. Such a
representation divides the phase space into sectors, each associated
with a symbol, in which each density has a definite
increasing/decreasing behavior.  The sectors' boundaries are the {\em
  nullclines}, i.e. the manifolds satisfying $g_i(x_1,x_2\ldots
x_N)=0$.  Our goal is to determine the conditions under which the
trajectory can cross a nullcline. This requires a density to change
from increasing to decreasing (or vice versa) and is equivalent to
determining when the density can have a maximum or minimum.

For example, a minimum for the variable $x_i$ corresponds to a
crossing of the nullcline $g_i=0$ from the region $g_i<0$ to the
region $g_i>0$. This is possible only if, somewhere on the nullcline,
the scalar product between the vector field $\vec{g}$ and the vector
$\nabla g_i$ (which is normal to the nullcline $g_i=0$) is positive:
\begin{equation}
\sum_{j\neq i} g_j(x_1,x_2\ldots x_N)\ \partial_{x_j} g_i(x_1,x_2,\ldots x_N)>0.
\end{equation}

The $i=j$ term is excluded since it is zero on the nullcline.  By
assumption, all the derivatives have fixed signs, and in any given
sector the $g_j$'s also have fixed signs (encoded in the associated
symbol).  If the symbol and derivative signs are such that each term
is negative, then the sum cannot be positive.
This implies the rule:\\
{\em A variable cannot have a minimum if all its repressors are
  increasing and all its activators are decreasing.}
%In the same way, one can derive the condition for maxima:\\

Similarly, for maxima:\\
{\em A variable cannot have a maximum if all its
  repressors are decreasing and all its activators are increasing.}

\begin{figure}[tb]
a)\hfill ~\\
 \epsfig{width = 0.4\textwidth,file=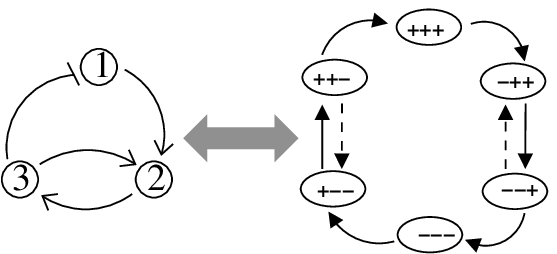}\\
b)\hfill c)\hfill ~\\
\hspace*{-0.5cm} \epsfig{width = 0.25\textwidth,file=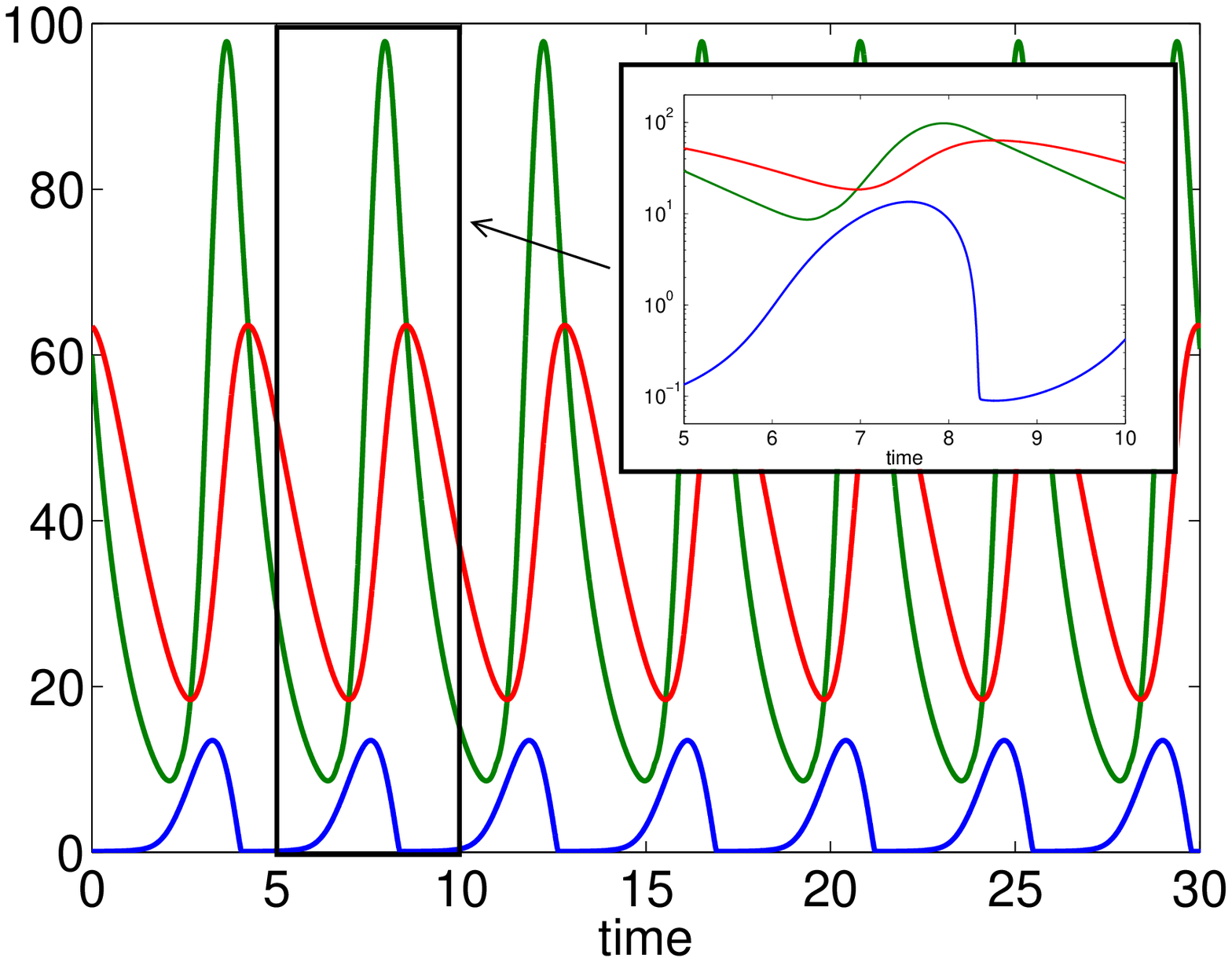}
\hspace*{-0.4cm} \epsfig{width = 0.25\textwidth,file=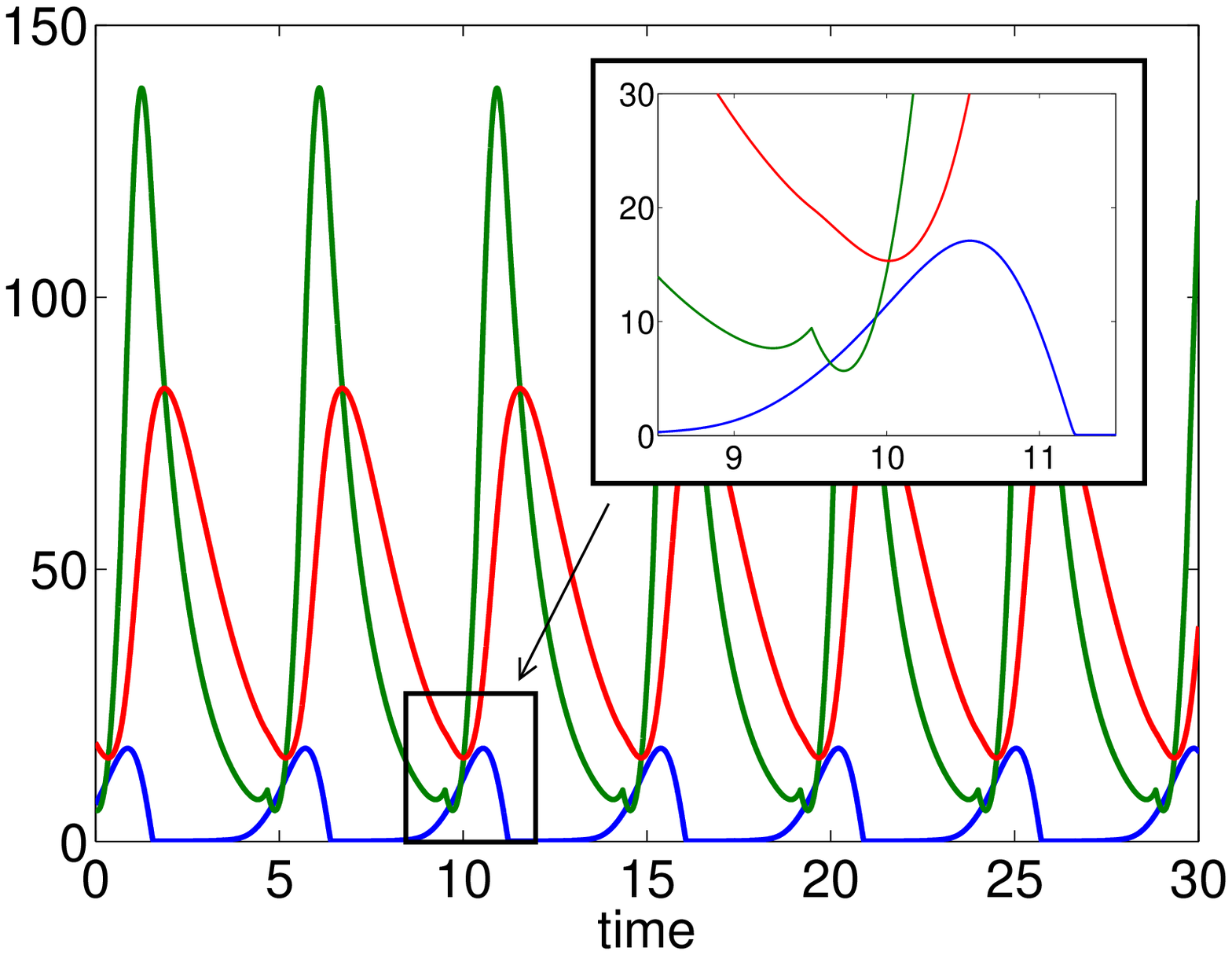}
\caption{Simple example of the use of symbolic dynamics.  (a) (left)
  Scheme of the network. We represent activation by a normal arrow,
  and repression by a barred arrow. (right) Corresponding transition
  network; we removed for clarity the symbols (+-+) and (-+-) which
  have no incoming links.  (b) Dynamics of the 3 variables as a
  function of time with a weak cross-link ($a=10$, see text), showing
  the transition cycle in solid arrows in (a). Inset shows the same on
  a log-scale. (c) Dynamics for a stronger cross-link ($a=50$, see
  text) which includes the transitions shown by dashed arrows, zoomed
  in the inset. In all plots $x_1$ is blue, $x_2$ is green and $x_3$
  is red.}
\end{figure}

Using the above two rules we can construct a network of allowed
transitions for a given biological module, with one node for each
symbol and an arrow for each transition that does not violate the
above rules.  Note that the transition network will only have arrows
connecting symbols differing by a single sign, because each maximum or
minimum corresponds to a single sign flip.  

We first consider the example network of Fig. 1a(left): a
three-species negative feedback loop with a cross-link from node 3 to
2 that introduces a positive feedback. By checking all the allowed
transitions we construct the corresponding transition network, shown
in Fig. 1a(right). For example, from the symbol $(-++)$ the transition
$(-++)\rightarrow(-+-)$ is ruled out because all the activators of
node 3 are increasing, therefore it cannot have a maximum.  Similarly
we rule out all transitions from it except $(-++)\rightarrow(--+)$.
The result, in this case, is a simple modification of the transition
network for a single negative feedback loop shown by the solid arrows
in Fig. 1a(right)\cite{patterns}.  With the cross link present, the
additionally allowed transitions are the ones shown with dashed
arrows.  The following dynamical system illustrates these
possibilities: $\dot{x_1}=c-x_1x_3/(k_1+x_1);
\dot{x_2}=x_1^2+a\left[\theta(x_3-k_2)-1\right]-x_2;
\dot{x_3}=x_2-x_3$.  The major nonlinearity is the Heaviside step
function: $\theta(x)=0$ for $x<0$, and $\theta(x)=1$ for
$x>0$\footnote{We can safely introduce a discontinuous field because
our argument works as long as trajectories are continuous.}. By
choosing parameters such that the cross link is weak ($c=30, a=10,
k_1=0.1, k_2=20$) one obtains dynamics of Fig. 1b, which is identical
to the simple 3-species loop.  As the strength of the cross link is
increased ($a=50$), the symbolic dynamics changes to also exhibit the
dashed transitions.  This is shown in Fig. 1c where variable $x_2$
develops a new small maximum, thus changing the symbolic dynamics.

\begin{figure}
a)\hfill ~\\
 \epsfig{width = 0.3\textwidth,file=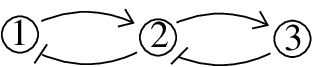}\\
b)\hfill~\\
 \epsfig{width = 0.5\textwidth,file=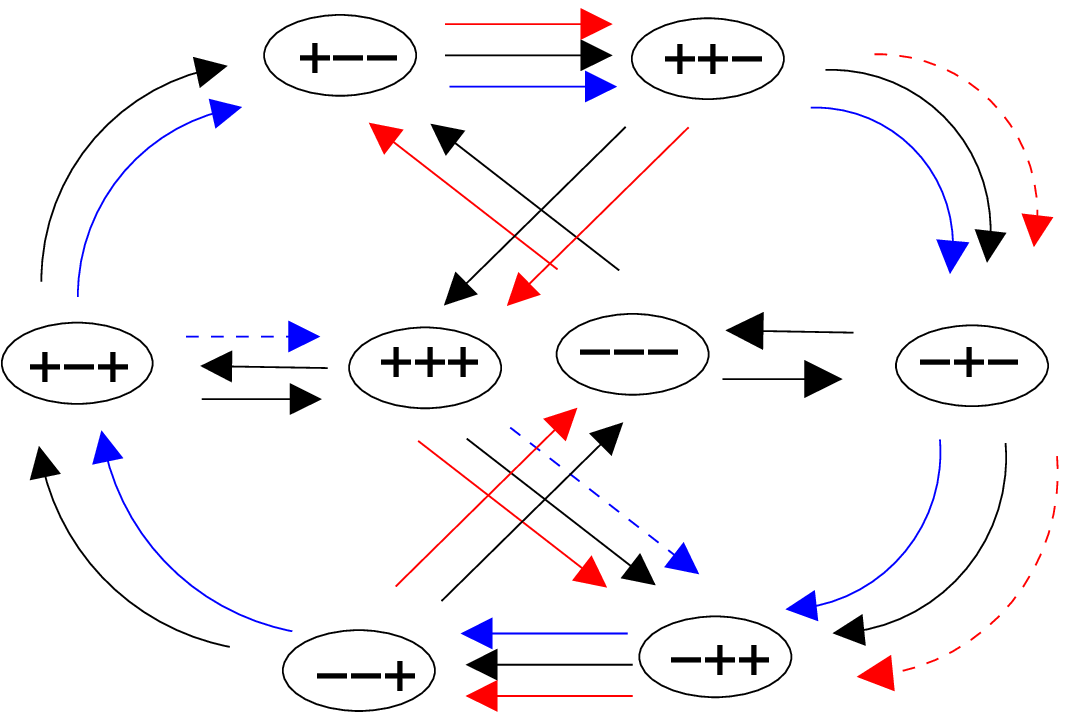}\\
 \caption{Network of two coupled two-species oscillators. (a)
   Structure of the network. (b) The transition network for this
   3-node system. Black arrows indicate all the allowed
   transitions. Blue arrows are the transitions actually observed in
   the HP system and red arrows are the transitions observed in the
   BHS model (see text). Dashed arrows indicate
   ``kicks", i.e., transitions which are {\em not} observed close to
   the Hopf bifurcation.}
\label{threelevels}
\end{figure}

We now move on to a system that exhibits a richer range of dynamical
behaviours (see Fig. \ref{threelevels}a). It consists of two negative
feedback loops, coupled via a shared species.  This network has been
widely studied in the ecological literature
\cite{blausius,hastings,stonehe} as a model for three trophic level
ecosystems: species $x_3$ feeds on $x_2$, and $x_2$ feeds on $x_1$.
The chaotic properties of this motif have been used to interpret data
from the Canadian lynx-hare cycle, showing irregular oscillations
\cite{gamarra}.

\begin{figure}[htb]
\includegraphics[width=8.5cm]{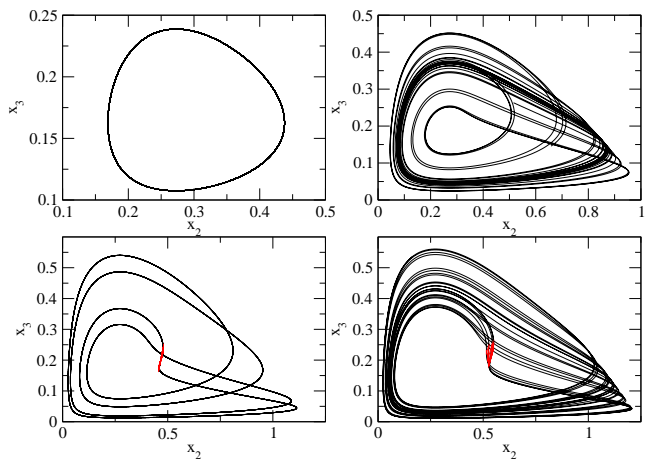}
\caption{Two dimensional projection of the attractor of the system of
equation (\ref{hastsystem}) for different values of the control
parameter $r=2.0$ (top-left), $r=2.6$ (top-right), $r=3.0$
(bottom-left) $r=3.3$ (bottom-right).}\label{figattr}
\end{figure}

\begin{figure}[htb]
\includegraphics[width=8.5cm]{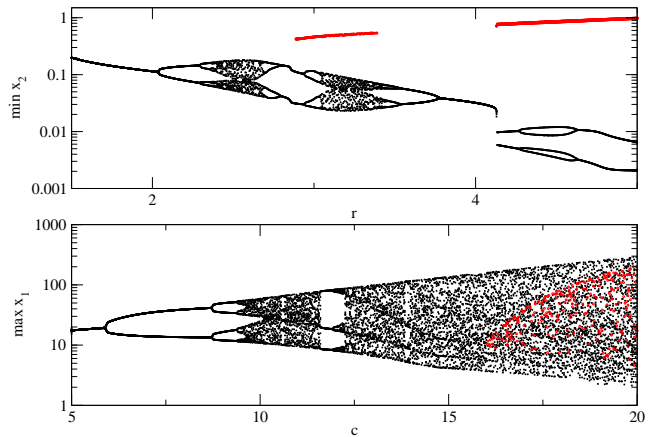}
\caption{Bifurcation diagrams. (top) HP model (\ref{hastsystem}), 
  minima of $x_2$ plotted versus $r$. (bottom) BHS model
  (\ref{sysblausius}), maxima of $x_1$ versus c.  In both plots, red
  dots indicate the appearance of ``kicks'' in the trajectory and
  symbolic dynamics (see text).  }
\label{figbifur}
\end{figure}

We consider first the Hastings-Powell (HP) model \cite{hastings} as a
dynamical system corresponding to this network:
\begin{eqnarray}\label{hastsystem}
\dot{x_1}&=&rx_1(1-kx_1)-\alpha_1x_1x_2/(1+b_1x_1)\nonumber\\
\dot{x_2}&=&-d_1 x_2+\alpha_1x_1x_2/(1+b_1x_1)-\alpha_2x_2x_3/(1+b_2x_2)\nonumber\\
\dot{x_3}&=&-d_2 x_3+\alpha_2x_2x_3/(1+b_2x_2)
\end{eqnarray}
with the following parameter choices: $\alpha_1=\alpha_2=4$,
$b_1=b_2=3$, $d_1=.4$, $d_2=.6$, $k=1.5$. By increasing the parameter
$r$, a stable limit cycle undergoes a series of period doubling
bifurcations, followed by the onset of chaos. A projection of the
attractor on the $x_2-x_3$ plane is shown in Fig.(\ref{figattr}).  The
chaotic trajectory looks similar to the periodic one, except for the
irregular behavior of the amplitude \cite{stonehe}.  This means that
the same sequence corresponding to the periodic orbit is observed
after the onset of chaos.  By increasing $r$ even more, we found a
regular window with a change in the symbolic dynamics (the ``kick'',
shown in red in the attractor in Fig. \ref{figattr} and in the
bifurcation diagram, Fig. \ref{figbifur}a, and corresponding to the
blue dashed transition in Fig. 2b). The kick is still present when, by
further increasing $r$, the dynamics becomes chaotic again.

The conclusion is that the same symbolic dynamics observed close to
the Hopf bifurcation is found in a large region of parameter space.
We compare the results with a different system corresponding to the
same network, the model by Blausius, Huppert and Stone
(BHS)\cite{blausius}:
\begin{eqnarray}\label{sysblausius}
\dot{x_1}&=&x_1-\alpha_1x_1x_2/(1+k x_1)\nonumber\\
\dot{x_2}&=&-d x_2+\alpha_1x_1x_2/(1+k x_1)-\alpha_2 x_2x_3\nonumber\\
\dot{x_3}&=&c(x_3^*-x_3)+\alpha_2 x_2x_3
\end{eqnarray}
with parameters $\alpha_1=2$, $\alpha_2=d=1$, $k=0.12$, $x_3^*=0.006$.
Here, a convenient control parameter is $c$.  We observe
the same scenario in the bifurcation diagram (see Fig.
(\ref{figbifur}b)): periodic orbit, then chaotic but same periodic
symbolic dynamics, then different symbolic dynamics in a regular
window and, finally, chaotic symbolic dynamics. Note, however, that
the periodic symbolic dynamics observed close to the Hopf bifurcation
is {\em different} from that observed in the HP model. 

To test the robustness of the two sequences, we tried to change the
functional form of the interaction between $x_2$ and $x_3$ by setting
$b_2=0$ in the HP model or, conversely, introducing saturated response
in the BHS model. We also tried to vary the parameters of both
systems, by up to 50\% from their default values. The two symbolic
sequences were not affected by any of these changes.  A possible cause
for this robust difference could be the logistic term in the first
equation of (\ref{hastsystem}), acting as a regulator so that the full
dynamics is bottom-up controlled.

The difference between the symbolic dynamics of the HP and BHS systems
can be used for model selection: in the example of the Canadian lynx
system, one has access only to the lynx population time series, but
temporal measurements of the hare and grass abundances could be used
to understand which model is more appropriate. Interestingly, from the
point of view of maxima/minima order, these two systems behave like
two different, single negative feedback loops \cite{patterns}: $3
\dashv 2\dashv 1 \dashv 3$ (HP system) and $1 \rightarrow 2
\rightarrow 3 \dashv 1$ (BHS system). Both these ``effective'' loops
would include an ``effective'' interaction between variables $x_1$ and
$x_3$.

So far we have gone from a known network to the transition network to
the time series. The reverse process uses the transitions observed in
an experimental time series to infer information about the underlying
network.  For example, the circadian oscillations of the three genes
$kaiA,B,C$ in cyanobacteria \cite{kucho} were shown in ref.
\cite{patterns} to have the following symbolic dynamics $(B,A,C)$:
$(+++)\rightarrow (-++)\rightarrow (--+)\rightarrow (---)\rightarrow
(+--)\rightarrow (++-)\rightarrow (+++)$.  Several networks are
consistent with this pattern -- the simplest is the loop $B\rightarrow
A\rightarrow C\dashv B$, as suggested in ref.  \cite{patterns}.
Experiments have shown that $A\rightarrow C$ and $C\dashv B$, and that
all three genes are essential for oscillations \cite{ishiura}.  With
this information we can get non-trivial guidelines for which further
interactions to look for experimentally: (i) $kaiA$ must be either
activated by $kaiB$, or repressed by $kaiC$ (or both), (ii) if $kaiA$
is not activated by $kaiB$ then, in addition to $kaiC\dashv kaiA$,
$kaiB$ must activate $kaiC$, so that the underlying network looks
similar to Fig. 2a. Of course, these predictions are for ``effective"
interactions, which at the molecular level could involve multiple
intermediates, such as chemical complexes and various protein activity
states. Ref. \cite{patterns} shows how the method can
reconstruct effective interactions even in the presence
of intermediate species.

This circadian example also points out how much information our method
provides.  The transition $(+++)\rightarrow(-++)$ means that either
$B\dashv A$ or $C\dashv A$. Later transitions show that either
$A\rightarrow B$ or $C\dashv B$, and $A\rightarrow C$ or $B\rightarrow
C$.  Even without the extra experimental information, our method
reduces the possibilities for the adjacency matrix of the underlying
network from $3^6$ to $5^3$, a factor of $\approx 6$.  In a general
$N$ node system, with $M$ independent observed transitions, the
fraction of allowed adjacency matrices is $[1-(2/3)^{N-1}]^M$; the
smaller the network and the more the transitions seen, the more useful
the method. A full oscillation cycle would show at least $N$
independent transitions.  If the system instead reaches a fixed point,
the transient can still be used.

Our method can be considered as complementary to the ``threshold
method'', described in Refs.
\cite{glass1,glass3,glasskauf,glasschemphys}, which provides a
different way of dividing the phase space into sectors, based on a
choice of thresholds for each variable. The ``threshold" method
generates a transition diagram, which depends on parameter values.  It
is particularly suited to cases where the input functions are Boolean
or step-like, so thresholds can be easily identified, and
self-interactions are piecewise linear \cite{plahte}.  Our
``derivative" method, in contrast, requires no choice of thresholds,
generates a diagram independent of parameter values, and
works for arbitrary self-interactions, but requires monotonicity in
the other interactions.

In summary, we showed a method to construct a symbolic transition
network that imposes a strong constraint on the dynamics monotone
systems, like many biological modules.  In all the cases we studied,
the periodic symbolic dynamics observed close to the Hopf bifurcation
is found in a large region of parameter space, even when the system
becomes chaotic. This explains the commonly observed phenomenology of
a chaotic attractor consisting of oscillations with randomly varying
amplitude \cite{stonehe}.  The oscillatory systems we looked at
produce a symbolic sequence identical to that of a single negative
feedback loop for most studied parameter values. By identifying these
loops, our method can be used to derive minimal models of complex,
oscillatory biological systems.

\begin{acknowledgments}
  We acknowledge Leon Glass for stimulating discussions and
  the Danish National Research Foundation and VILLUM KANN RASMUSSEN
  Foundation for funding.
\end{acknowledgments}

\end{document}